\documentclass{iopart}
\usepackage{graphicx}

\begin{document}

\title[Driving propagation of ferromagnetic $\pi$/2 domain walls in nanostripes]{Driving 
large-velocity propagation of ferromagnetic $\pi$/2 domain walls in nanostripes of cubic-anisotropy materials}

\author{Andrzej Janutka\dag, Przemys{\l}aw Gawro\'nski\ddag\hspace{1ex}and Pawe{\l} S Rusza{\l}a\S}

\address{\dag\hspace{0.5ex}Department of Theoretical Physics, Wroclaw University of Technology, 50-370 Wroc{\l}aw, Poland} 

\address{\ddag\hspace{0.5ex}Faculty of Physics and Applied Computer Science, AGH University of Science and Technology, 30-059 Krakow, Poland} 

\address{\S\hspace{0.5ex}Faculty of Fundamental Problems of Technology, Wroclaw University of Technology, 50-370 Wroc{\l}aw, Poland}

\begin{abstract}
We study the externally-driven motion of the domain walls (DWs) of the $\pi/2$ type 
 in (in-the-plane ordered) nanostripes with the crystalline cubic anisotropy.
 Such DWs are much narrower than the transverse and vortex $\pi$ DWs in the soft-magnetic
 nanostripes while they propagate much faster, thus, enabling dense packing of magnetization 
 domains and high speed processing of the many-domain states. The viscous current-driven
 motion of the DW with the velocity above 1000m/s under the electric current
 of the density $\sim 10^{12}$A/m$^{2}$ is predicted to take place in the nanostripes
 of the magnetite. Also, the viscous motion with the velocity above 700m/s
 can be driven by the magnetic field according to our solution to a 1D analytical 
 model and the micromagnetc simulations. Such huge velocities are achievable 
 in the nanostripes of very small cross-sections (only 100nm width and 10nm thickness). 
 The fully stress driven propagation of the DW in the nanostripes of cubic magnetostrictive
 materials is predicted as well. The strength of the DW pinning to the stripe notches and
 the thermal stability of the magnetization during the current flow are addressed.
\end{abstract}


\ead{Andrzej.Janutka@pwr.edu.pl}
\maketitle

\section{Introduction}

The domain walls (DWs) in the ferromagnetic nanowires (nanostripes) are
 the objects of a huge current interest because of concepts of miniaturization
 of the non-volatile memory and of the magnetics-based logic. Especially, the idea
 of the 3D digital recording known as the "domain-wall racetrack memory" has 
 taken enormous attention of the magnetics community \cite{par08,all05}.
 It is based on positioning a train of the magnetic DWs in the nanowire 
 with the electric current. However, the racetrack concept
 has met difficulties. The most challenging of them is claimed
 to be a rapid increase of the temperature during the current-driven propagation
 of the DWs (a fast Joule heating) \cite{tho11,zha14}. The minimum value of the spin
 transfer torque (STT), thus, the minimum intensity of the applied current and the minimum
 pace of the Jolue heating are limited by two factors. These are the DW pinning to natural
 and artificial (position-stabilizing) defects of the stripe and the need for 
 reducing the operating time via increasing the DW velocity. Overcoming this difficulty
 has motivated the evolution of the DW racetrack architecture whose primary step was usage of the 
 nanostripes with perpendicular magnetic anisotropy (PMA) instead
 of the longitudinally magnetized stripes \cite{par15}. However, since the PMA
 stripes are the uniaxial ferromagnets in principle, the field and current driven
 motions of the DW are turbulent whereas the record stability requires them to be
 viscous \cite{met07,mar12}. By the viscous motion we mean a uniform translational movement
 without the magnetization precession in the transverse plane (it takes place below
 the Walker breakdown).
 This requirement is satisfied in double-layer (or triple-layer) nanowires composed
 of a PMA ferromagnet (ferromagnets) and a non-magnetic metal. The stable (uniform) and fast
 current-driven propagation of the DW stack is possible utilizing the spin-Hall effect in the
 non-magnetic layer provided
 the global chirality of the DW system is stabilized by an interface-induced 
 (Dzyaloshinskii-Moriya-like) anisotropy \cite{emo13}.
 
Simplifying the racetrack architecture is of importance with
 regard to the purpose of manufacturing the 3D memory device.
 The challenge is the stabilization of the DW structure
 in a simple nanowire (a single-layer nanostripe), thus, shifting the Walker-breakdown
 point upward the axis of the magnetic-field intensity. Our hypothesis is that the presence 
 of two easy (or two hard) anisotropy axes could improve the stability. In particular,
 the racetrack created of a ferromagnetic material with a multi-axis crystalline anisotropy
 is an option to consider. Small width and high mobility of the DWs are desired and these
 advantages determine our preferences in terms of the material and sizes of the nanostripe. 

In a recent paper, analyzing the structure of the ferromagnetic DWs in the nanostripes
 made of cubic anisotropy materials, we (A.J. and P.G.) have predicted the $\pi$/2 DW
 to stabilize in sufficiently-thin stripes (e.g. of Fe$_{3}$O$_{4}$) provided the stripe
 axes coincide with the hard directions of the crystalline anisotropy \cite{jan14}.
 Those DWs separate the domains magnetized along the easy axes in the stripe plane
 (diagonal to the stripe axes). Such an ordering requires the saturation magnetization
 to be small enough that the magnetostatic (shape-anisotropy) energy 
 $E_{ms}=-\mu_{0}\int\bi{m}\cdot\bi{H}_{d}(\bi{m}){\rm d}V$ is comparable
 to the exchange energy $E_{ex}=A\int\sum_{i=x,y,z}|\nabla m_{i}|^{2}{\rm d}V$
 and it does not exceed the energy of the crystalline anisotropy
 $E_{an}\approx K_{1}/M^{4}\int m_{x}^{2}m_{y}^{2}{\rm d}V$ (writen up to the leading term).
 Here $\bi{m}$ denotes the magnetization, ($M\equiv|\bi{m}|$ and $z$ axis is normal to the stripe 
 plane), $\bi{H}_{d}$ denotes the dipole field, $A$ and $K_{1}<0$ are the exchange stiffness 
 and cubic-anisotropy constant, respectively. In the static case, the cubic anisotropy is
 equivalent to the four-fold in-the-plane anisotropy due to the planar alignment of
 the magnetization.
 The width of the $\pi$/2 DWs has appeared to be independent of the stripe width
 and much smaller than the width of the transverse or vortex DWs in soft-ferromagnetic 
 nanostripes or in thicker nanostripes of the cubic materials as well. 
 The small width of the $\pi$/2 DWs is expected to enable packing the bits
 (domains) in the racetrack with a density comparable to that in the PMA nanostripes.

The aim of the present paper is to study the field-driven, current-driven, and stress-driven
 dynamics of the head-to-head (tail-to-tail) $\pi$/2 DWs in the nanostripes made
 of cubic-anisotropy ferromagnet with regard to the racetrack design.
 We estimate the velocities of the DW under different 
 external drivings. The strength of pinning of the $\pi$/2 DW to a notch at one of 
 the stripe edges is examined and the depinning current is determined. 
 For the minimum current intensity established, the pace of the temperature increase 
 due to the Joule heating is evaluated. Hence, we determine the main limitations of 
 the $\pi$/2-DW based racetracks.

The recent rapidly increasing interest in the methods of control of the magnetization 
 state via the stress application (the straintronics) is motivated by the hope
 for reducing the energy consumption compared to the application of the magnetic field
 or the electric current. The strong crystalline anisotropy, in particular, the cubic
 anisotropy, is often accompanied by a significant magnetostriction \cite{kro03}.
 This is shown below to offer a possibility of driving the head-to-head
 (tail-to-tail) $\pi$/2 DW propagation
 via the application of the mechanical stress. The study of the efficiency of such
 a driven motion completes the present analysis of the potential of the $\pi$/2 DW
 for the racetracks.

In sec. II, we provide a background of the $\pi$/2 DWs in the cubic-anisotropy
 materials and consider the ways of driving the DW propagation.
 The field and current driven motions of the 
 $\pi$/2 DWs in the nanostripe are investigated in subsec. IIIa, while the depinning 
 from the notch is studied in subsec. IIIb. Sec. IV is devoted to analyzing
 the stress-driven motion of the $\pi$/2 DW. Conclusions are collected in sec. V.

\section{Background}

\begin{figure}
\unitlength 1mm
\begin{center}
\begin{picture}(175,30)
\put(25,-5){\resizebox{87mm}{!}{\includegraphics{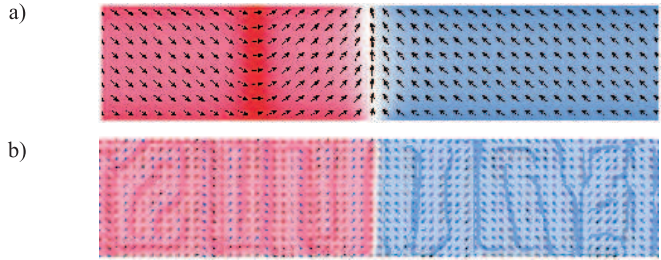}}}
\end{picture}
\end{center}
\caption{A static $\pi$/2 head-to-head DW in the neighborhood of 
a head-to-tail $\pi$/2 DW in: a $w\times\tau=400\times10$nm$^{2}$ cross-section
nanostripe of the magnetite (a), and a static $\pi$/2 head-to-head DW in the neighborhood of 
many head-to-tail DWs in a $w\times\tau=400\times40$nm$^{2}$ cross-section nanostripe of the Terfenol-D (b). The colors and their intensity indicate the sign and value of the projection of the magnetization on the long axis of the stripe.}
\end{figure}

In the previous work, we have performed the micromagnetic simulations of the DW formation 
 in the nanostripes made of cubic-anisotropy materials and we have developed an analytical
 description of the DW structures \cite{jan14}. The density of the crystalline-anisotropy
 energy of the relevant cubic ferromagnets takes the form 
\begin{eqnarray}
{\cal H}_{a}=\frac{K_{1}}{M^{4}}\left(m_{x}^{2}m_{y}^{2}+m_{y}^{2}m_{z}^{2}+m_{y}^{2}m_{z}^{2}\right)
+\frac{K_{2}}{M^{6}}m_{x}^{2}m_{y}^{2}m_{z}^{2},
\end{eqnarray}
see e.g. \cite{kro03}. For a certain range of the material parameters (when the exchange
 lengths $l_{ex}=\sqrt{2A/\mu_{0}M^{2}}$, $l_{K}=\sqrt{-A/K_{1}}$ are comparable), in very thin
 nanostripes [of the thickness of the order of $l_{ex}^{2}/l_{K}$)], the $\pi$/2 DWs are
 stable solutions to the Landau-Lifshitz-Gilbert equation. In particular, they have been 
 found with the material constants of the magnetite (Fe$_{3}$O$_{4}$) for the nanostripe  
 thicknesses of up to 20nm and with the parameters of the Terfenol-D (Tb$_{0.3}$Dy$_{0.7}$Fe$_{2}$)
 for the nanostripe thicknesses of up to 50nm, in the case of the crystallographic axes aligned
 parallel to the relevant stripe axes. The following material parameters of Fe$_{3}$O$_{4}$ have
 been included: $M=3\cdot10^{5}$A/m, $K_{1}=-1.1\cdot10^{4}$J/m$^{3}$,
 $K_{2}=-0.3\cdot10^{4}$J/m$^{3}$, $A=1.2\cdot10^{-11}$J/m,
 the Gilbert damping constant
 $\alpha=0.014$, and the gyromagnetic ratio $\gamma\approx\mu_{0}^{-1}\cdot2.21\cdot10^{5}$m/As,
 \cite{gal54,boz55,bra08,nag14}. We remark that the effective saturation magnetization for
 the film of the magnetite is significantly smaller than for the bulk material
 ($M=4.8\cdot10^{5}$A/m), which is due to the presence of anti-phase areas \cite{bra08}.
 The mechanism of the anti-phase appearance is connected to an antiferromagnetic interaction
 at the anti-phase boundaries \cite{sof11}. It is local and not expected to influence the energy
 of the single-ion anisotropy, thus, we use the anisotropy constants of the bulk.
 Let us mention that, contrary to previous reports on the attenuation of the anisotropy 
 in the Fe$_{3}$O$_{4}$ films, careful analysis suggests the four-fold in-the-plane anisotropy
 of the thin layers of the magnetite to be even stronger than in the thick films (the bulk
 limit) \cite{sch15}.  
 The relevant parameters of Tb$_{0.3}$Dy$_{0.7}$Fe$_{2}$ are following: $M=7.5\cdot10^{5}$A/m,
 $K_{1}=-8.7\cdot10^{5}$J/m$^{3}$, $K_{2}=2.3\cdot10^{6}$J/m$^{3}$,
 $A=9.0\cdot10^{-12}$J/m, $\alpha=0.1$, \cite{del04,fas11}.

The presence of the $\pi$/2 DWs is a consequence of a large in-the-plane tilt of the direction
 of the domain magnetization from the long axis of the stripe, (see figure 1),
 that results from the strong crystalline anisotropy
 (in the case $K_{1}<0$). The competition between the crystalline and shape anisotropies
 (the effect of the demagnetization field) results in a limitation on the length of the uniformly
 magnetized domains, thus, it leads to the creation of periodic structures of head-to-tail $\pi$/2
 DWs in the nanostripe (modulated superdomains). The superdomains of different projections
 of the magnetization onto the long axis of the stripe are separated by head-to-head
 or tail-to-tail $\pi$/2 DWs, (similar structures of $\pi$/2 DWs are observed in 
 nm-thickness rings of Fe$_{3}$O$_{4}$, \cite{fon11}).
 Hence, unlike in the proposals of the $\pi$-DW racetrack, we consider
 a record of bits which are written in patterned superdomains. The period of the patterns
 depends on the saturation magnetization (on the ratio $\mu_{0}M^{2}/K_{1}$, where $\mu_{0}$
 denotes the vacuum permeability) and on the thickness
 of the stripe since these two factors influence the demagnetization (dipole) field. 

Similar to case of the $\pi$ DWs in the stripes and wires, the uniform electric current through
 the cubic-anisotropy nanostripe induces STT
 in the DWs driving the propagation of all $\pi$/2 DWs through the nanostripe
 in the same direction. Also, similar to the case of the $\pi$ DWs, the application
 of an external longitudinal magnetic field drives the neighboring head-to-head and tail-to-tail 
 $\pi$/2 DWs to propagate in the opposite directions because the superdomains magnetized
 parallel (antiparallel) to the field must expand (shrink) \cite{bea07}. 

Another way of driving the motion of the DW can arise when the 
 neighboring ferromagnetic domains are not magnetized parallel to each other as it is in the 
 nanostripes under consideration. Any uniform in-the-plane stress applied in a different direction
 than the stripe axes is expected to induce a difference
 of the anisotropy energies of the neighboring domains due to the magnetostriction, thus,
 to drive expansion or shrinking of the domains \cite{bri10,par12}. In the case of a pair
 of one head-to-head and one tail-to-tail DWs magnetized in the same direction (of the opposite
 chiralities), the application of such a stress is expected to result in the mutual collision
 of the DWs and, provided the DW motion was viscous, in their annihilation.
 Hence, the stress application is a potential alternative to usage of the external magnetic field,
 in particular, with the purpose of manipulating the bit record via the annihilation
 of the DW pairs (the reversal of the superdomain magnetization). 

Because of the above, materials of strong (giant) magnetostriction are of our especial interest.  
 Within the class of them, besides the magnetite and Terfenol-D, we have managed to simulate
 the $\pi$/2 DWs in thin nanostripes of the cobalt ferrite (CoFe$_{2}$O$_{4}$). Remark
 that the nanotechnology of the Terfenol-D remains a challenge however.
 Unfortunately, simulating the domain wall creation in the nanostripes of the cubic-anisotropy
 and high-magnetostriction Fe$_{1-x}$Ga$_{x}$ (Galfenol-like) alloys (of the 10nm thickness
 at least), we have not managed to find the $\pi$/2 DWs for any concentration $x$.
 This fact resembles quite high ratio of the exchange length of the crystalline 
 anisotropy to the magnetostatic exchange length for Fe$_{1-x}$Ga$_{x}$, \cite{jan14}. 

The density of the magnetoelastic energy of the cubic ferromagnet is of the form
\begin{eqnarray}
\fl
{\cal H}_{me}=-\frac{3}{2M^{2}|{\bf\sigma}|}\lambda_{100}\left(m_{x}^{2}\sigma_{x}^{2}+m_{y}^{2}\sigma_{y}^{2}+m_{z}^{2}\sigma_{z}^{2}\right)\nonumber\\
-\frac{3}{M^{2}|{\bf\sigma}|}\lambda_{111}\left(m_{x}m_{y}\sigma_{x}\sigma_{y}
+m_{x}m_{z}\sigma_{x}\sigma_{z}+m_{y}m_{z}\sigma_{y}\sigma_{z}\right),
\end{eqnarray}
where $\bi{\sigma}$ denotes the stress vector, and $M\equiv|\bi{m}|$ is the saturation
 magnetization \cite{kro03}. Applied to the cubic media, one uses
 the name 'giant magnetostriction' referring 
 to the materials of large value of the volume magnetostriction parameter;
 $\lambda=2/5\lambda_{100}+3/5\lambda_{111}$, where $\lambda_{100}$, $\lambda_{111}$ 
 determine the tetragonal and rhombohedral magnetostrictions respectively.
 The last formula holds for the polycrystalline ferromagnets \cite{gro14}.
 When inducing the DW translation in a single-crystal film using the stress, the strong
 rhombohedral magnetostriction is needed. The energy of the tetragonal magnetostriction
 of the neighboring
 domains is independent on whether they are magnetized parallel or not provided the absolute
 values of the angles between the domain magnetization and the stripe axes are globally conserved.
 The energy of the rhombohedral magnetostriction makes a difference. 
 Moreover, large value of the ratio $\lambda_{111}/\lambda_{100}$ is an advantage (see Appendix B).
 Therefore, the materials of strong rhombohedral magnetostriction: Fe$_{3}$O$_{4}$,   
 ($\lambda_{100}=-19\cdot10^{-6}$ and $\lambda_{111}=81\cdot10^{-6}$);
 CoFe$_{2}$O$_{4}$; ($\lambda_{100}=-590\cdot10^{-6}$ and $\lambda_{111}=120\cdot10^{-6}$) \cite{boz55}; 
 Tb$_{0.3}$Dy$_{0.7}$Fe$_{2}$; ($\lambda_{100}=-90\cdot10^{-6}$
 and $\lambda_{111}=-1350\cdot10^{-6}$) \cite{del04}; are of our especial interest. We notice that, in the contrast
 to them, the giant-magnetostriction Galfenol is of a near-zero rhombohedral magnetostriction. 

\section{Field and current-driven domain-wall motion}

\subsection{DW mobility}

\begin{figure}
\unitlength 1mm
\begin{center}
\begin{picture}(175,62)
\put(25,-5){\resizebox{87mm}{!}{\includegraphics{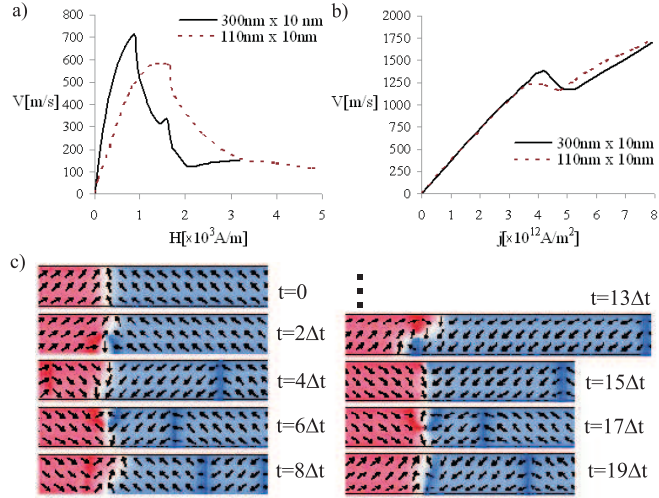}}}
\end{picture}
\end{center}
\caption{Top: the dependence of the velocity 
of the $\pi$/2 head-to-head DW in the $w\times\tau=110\times10$nm$^{2}$ cross-section
and $w\times\tau=300\times10$nm$^{2}$ cross-section nanostripes of the magnetite
on the longitudinal magnetic field (a), and on the current intensity (b).
Bottom: dynamical transformations of the propagating head-to-head DW
in the $w\times\tau=300\times10$nm$^{2}$ cross-section nanostripe of the magnetite under the 
the longitudinal field $H=1670$A/m. The time step is $\Delta t=3.0$ns, the reference frame is 
connected to the DW moving right the picture.}
\end{figure}

In order to characterize the efficiency of the field-driven DW propagation and of the current-driven 
 DW propagation, we determine the relevant mobilities of the head-to-head $\pi$/2 DW with  
 micromagnetic simulations. In particular, the maximum
 velocity of the viscous motion and the field intensity (current density) of the Walker breakdown
 are the characteristics which limit the applicability of DW-based devices.
 The velocity-field and velocity-current plots are shown in figure 2a and figure 2b, respectively,
 for the Fe$_{3}$O$_{4}$ nanostripes of the widths $w$=110nm and $w$=300nm and of
 the thickness $\tau$=10nm. In the simulations (using the OOMMF package \cite{oommf}), the cell size
 of the grid discretization is 10nm, while the stripe length is 30$\mu$m. In the plots,
 values of $v(H)$ and $v(j)$ are average velocities over 15$\mu$m propagation distance. 
 It should be stressed that this propagation distance is sufficient to determine $v(H)$ in the 
 considered range of the magnetic field and to determine $v(j)$ in the regime of the viscous
 motion (below the critical current of the Walker breakdown). However, in the regime of turbulent
 motion (above the Walker breakdown), the propagation distance of 15$\mu$m
 is insufficient to evaluate $v(j)$ as an average over many cycles of DW transformations 
 since the relevant propagation time is shorter than the period of the dynamical
 transformations of the DW. Therefore, the plotted in figure 2b $v(j)$ curves do not resemble
 typical shape on the right-hand side of the local maximum
 (the Walker-breakdown peak) \cite{thi05,bea08}. On the other hand,
 for a reason discussed below (the fast Joule heating), there is no sense of simulating
 the DW motion on a wider time scale than the present one.  

For the purpose of analyzing the plotted data, a simple 1D model
 of the field and current induced motion of the head-to-head (tail-to-tail) $\pi$/2 DW
 is formulated in Appendix A. According to our analytical model, in the viscous-motion regime,
 the DW velocity depends on the intensity of the external field via the Walker-like formula 
 $v(H)=\gamma\mu_{0}\Delta H/\alpha$, where the DW width corresponds to the exchange length 
 of the crystalline anisotropy $\Delta\approx\sqrt{-A/K_{1}}$. Here, $\gamma$ denotes
 the gyromagnetic factor, $\mu_{0}$ the vacuum permeability, $\alpha$ the Gilbert damping
 constant, $A$ the exchange stiffness. Basing on a 2D model of reference \cite{jan14},
 the estimated in Appendix A width of the $\pi$/2 DW has been renormalized by the factor
 of two; $\Delta\approx2\sqrt{-A/K_{1}}$ which leads to $\Delta\approx 66$nm
 for Fe$_{3}$O$_{4}$, and the DW mobility $v(H)/H=\gamma\mu_{0}\Delta/\alpha=1.04$m$^{2}$/As. 
 This value coincides with the mobility determined from figure 2a for the $\pi/2$ DW 
 in the aspect-ratio $w/\tau=30$ nanostripe. For the nanostripe of $w/\tau=11$, 
 we have established $v(H)/H\approx0.55$m$^{2}$/As.
 Thus, the mobility of the $\pi/2$ DWs in the Fe$_{3}$O$_{4}$ nanostripes is found
 to be at least two times higher than estimated and measured in reference \cite{bea08} for the 
 vortex DWs in the Py nanostripes of the aspect ratio $w/\tau=10\div25$; 
 $v_{Py}(H)/H\approx0.2\div0.3$m$^{2}$/As, (here and below, the parameters of 
 the Py systems are indexed with the relevant superscript). Moreover, for the stripes
 of similar aspect ratio, the Walker breakdown fields in the magnetite structures are 
 larger than in the Py structures. In the consequence, the maximum velocity of the
 field-driven viscous motion of the $\pi/2$ DW significantly exceeds that 
 of the vortex DW in the Py nanostripes.

Similar to the $\pi$ DW in the Permalloy, the head-to-head $\pi/2$ DW undergoes
 a series of transformations into the vortex or/and antivortex DW above the Walker breakdown.
 There is characteristic peak in the field dependence of the DW velocity above
 the Walker breakdown that is well seen in $v(H)$ plot for the nanostripe of the 300nm width
 (centered at the field value $H\approx1.5\cdot10^{3}$A/m) in figure 2a.
 It is accompanied by a transition between regimes of oscillatory and chaotic
 dynamical transformations of the DW following the nomenclature of reference \cite{lee07},
 where similar peak has been found for $\pi$ DWs in Py nanostripes. 
 The transformation of the $\pi/2$ DW into the antivortex DW
 is accompanied by the creation of two head-to-tail $\pi/2$ DWs on both sides
 of the head-to-head DW [see the magnetization snapshots for $0\le t\le8\cdot 3{\rm ns}$
 in figure 2c], whereas, the  
 transformation into the vortex DW and backward into the $\pi/2$ DWs takes place during
 the passage of the head-to-head DW through a head-to-tail DW [see the magnetization
 snapshots for $8\cdot 3{\rm ns}\le t\le15\cdot 3{\rm ns}$ in figure 2c].  

Within the 1D model of Appendix A, the velocity of the current-driven
 viscous propagation of the head-to-head (tail-to-tail) $\pi/2$ DW is dependent
 of the current intensity following $v(j)=\beta\eta j/\alpha$. Here, $\eta$ is the constant
 of the STT strength; $\eta=Pg\mu_{B}/2eM$, where, $P$, $g$, $\mu_{B}$, and $e$ denote
 the spin polarization, the Lande factor, the Bohr magneton, and the electron charge,
 respectively \cite{bea08}. In the simulations of the Fe$_{3}$O$_{4}$ nanostripes; $P=0.94$,
 thus, $\eta=1.9\cdot10^{-10}$m$^{3}$/C, \cite{ces05}. 
 The ratio $\beta/\alpha$ determines the non-adiabaticity of the STT in the LLG
 equation \cite{zha04}. 
 Assuming the corresponding non-adiabaticity indicator in the LL equation to be equal
 to one, we take $\beta/\alpha=2$, (see \cite{tse08}). Let us mention that, using the 
 $\pi$ DWs in Py nanostripes (the polarization $P_{Py}=0.5$, the saturation magnetization
 $M_{Py}=8.6\cdot10^{5}$A/m), the relation $\beta_{Py}/\alpha_{Py}=2$ has been verified
 via measuring the DW velocity-to-current
 ratio $v_{Py}(j)/j=0.73\cdot10^{-10}$m$^{3}$/C, \cite{hay07}.

We emphasize that current-related mobility of the head-to-head $\pi/2$ DWs in
 the magnetite nanostripe; $v(j)/j$ is higher for the magnetite nanostripe than
 for the head-to-head $\pi$ DWs in the Py nanostripe, (because of higher spin polarization
 and lower saturation magnetization). Also, the breakdown current of the $\pi/2$ DW
 in the magnetite (the critical current of the transition between viscous and turbulent motion
 regimes) is at least twice the breakdown current for the vortex DW 
 in the Py nanostripe \cite{hay07}.
 Therefore, the velocity of the simulated current-driven propagation of the $\pi/2$ DW is
 as high as 1100m/s for the current density of $3\cdot 10^{12}$A/m$^{2}$. 
 It significantly exceeds the highest reported velocity of the DW in a sandwiched
 (triple-layer) nanostripe; 750m/s, that has been induced with similar value of the current
 density \cite{yan15}. However, unlike the sandwiched nanostripe of reference \cite{yan15}
 (an artificial antiferromagnet), our ferromagnetic system enables the field-driven propagation
 of the DW as well! Moreover, following the present simulations, under currents exceeding
 the breakdown value, on a certain time scale that is shorter than the time of the dynamical
 transformation of the DW, (thus, the DW motion is thought of to be almost viscous),
 the velocity of the head-to-head $\pi/2$ DWs can be even higher than 1100m/s.

\begin{figure}
\unitlength 1mm
\begin{center}
\begin{picture}(175,37)
\put(25,-5){\resizebox{87mm}{!}{\includegraphics{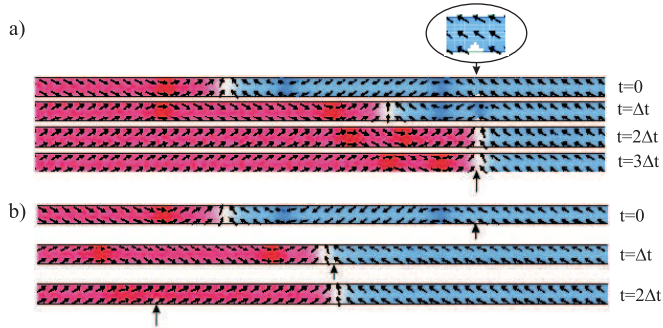}}}
\end{picture}
\end{center}
\caption{The current-driven motion of the $\pi$/2-DW in the presence of a pinning site at the 
lower edge of the Fe$_{3}$O$_{4}$ stripe of the $w\times\tau=110\times10$nm$^{2}$ cross-section.
The current densities are $j=1\cdot 10^{12}$A/m$^{2}$ (a), and $j=1.5\cdot 10^{12}$A/m$^{2}$ (b). The time step is $\Delta t=3.2$ns. In b) the reference frame is moving relative to the pinning
site.}
\end{figure}

The field-driven and current-driven motions of the head-to-head (tail-to-tail) $\pi/2$ DWs are
 different in another aspect than previously mentioned as well. Unlike the current, the
 longitudinal external field does not drive the head-to-tail $\pi/2$ DWs to move. It is
 because the propagation of the head-to-tail DW is not accompanied by any change of the Zeeman
 energy in itself while the Zeeman energy changes during shrinking or expanding the
 superdomains (the propagation of the head-to-head and tail-to-tail DWs).
 In the consequence, the head-to-head $\pi/2$ DW propagates pushing
 one or two head-to-tail DWs. Any next head-to-tail DW that appears on the way of the complex
 of the three $\pi/2$ DWs is annihilated during the collision, which is accompanied by
 a reduction of the complex to a pair of the DWs (one of the head-to-head type and one
 of the head-to-tail type). 

\subsection{DW depinning and issue of thermal stability}

\begin{figure}
\unitlength 1mm
\begin{center}
\begin{picture}(175,72)
\put(25,-5){\resizebox{87mm}{!}{\includegraphics{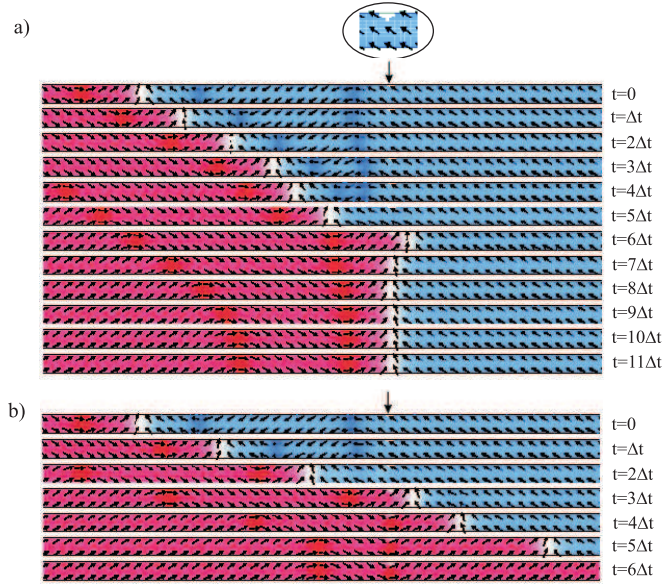}}}
\end{picture}
\end{center}
\caption{The current-driven motion of the $\pi$/2-DW in the presence of a pinning site at the 
upper edge of the Fe$_{3}$O$_{4}$ stripe of the $w\times\tau=110\times10$nm$^{2}$ cross-section. 
The current densities are $j=2.5\cdot 10^{11}$A/m$^{2}$
(a), and $j=5\cdot 10^{11}$A/m$^{2}$ (b). The time step is $\Delta t=3.2$ns.}
\end{figure} 

The issue to address is the strength of the pinning of the head-to-head (tail-to-tail) DW
 by stripe constrictions. The periodically distributed constrictions are considered
 to be pinning centers that stabilize the DW position in the racetrack
 counteracting mutual interactions between the DWs. We simulate the depinning 
 of the DW from an artificially created (at one of the stripe edges) notch. The depth of the
 notch is 3/10 of the stripe width while
 its shape is triangular. Remark that, in usual simulations of the current-driven motion
 of the DW, the depth of the notch is limited by the assumption
 of the uniform density of the current (the parameter of the STT is taken to be constant).
 That assumption could be contested when a variation of the stripe cross-section was large
 (the constriction of the stripe was very narrow).

According to figure 3 and figure 4, the courses of the pinning/depinning processes
 are dependent on the orientation of the pinning-site triangle relative to the directions 
 of the magnetization of the closest domains. The strongest pinning is present when the 
 the domains are magnetized parallel to the border of the notch, and this observation
 concerns the head-to-head (tail-to-tail) $\pi$/2 DWs as well as the head-to-tail $\pi$/2 DWs. 
 The density of the depinning current for the head-to-head $\pi$/2 DW in
 $w\times\tau=110$nm$\times10$nm magnetite stripe appeared
 not to exceed $1.5\cdot 10^{12}$A/m$^{2}$, which is several times higher value
 than achievable with PMA nanostripes while lower than the depinning current 
 of the transverse DWs in the Py nanostripes of similar cross sections with
 similar notches \cite{tho11}. Unfortunately, details of the notch shape are not standardized.
 A quite independent of those details indicator of the efficiency of the current-driven
 depinning is the ratio of the depinning current $j_{p}$ to the depinning field $H_{p}$.
 
For the given depth of the notch, we have confirmed the theoretical
 relation $j_{p}=\alpha\mu H_{p}/\beta\eta$ with the simulations. Here $\mu\equiv v(H)/H$
 denotes the DW mobility. For instance, the ratio
 $j_{p}/H_{p}=1.46\cdot10^{12}{\rm Am}^{-2}/955{\rm Am}^{-1}=1.5{\rm nm}^{-1}$
 has been obtained from the simulations of $w\times\tau=110{\rm nm}\times10{\rm nm}$ magnetite
 nanostripe with the geometry of figure 3 (the notch at the lower edge, the DW magnetized upward
 the stripe plane). It coincides with the theoretical value 
 $\alpha\mu/\beta\eta=0.55{\rm m}^{2}{\rm C}^{-1}/2\cdot1.9\cdot10^{-10}{\rm m}^{3}{\rm C}^{-1}=1.5{\rm nm}^{-1}$
 Following reference \cite{hay06}, the maximum of the relevant ratio for the 
 transverse DWs in the $w\times\tau=300{\rm nm}\times10{\rm nm}$ Py nanostripe takes the value of
 $j_{p}^{Py}/H_{p}^{Py}=3\cdot10^{12}{\rm Am}^{-2}/3500{\rm Am}^{-1}=0.86{\rm nm}^{-1}$. 
 Extrapolating this ratio to the nanostripe of $100$nm width (dividing it by three) leads
 to five times smaller value than for the head-to-head $\pi$/2 DW in the magnetite nanostripe
 ($j_{p}/H_{p}=5j_{p}^{Py}/H_{p}^{Py}$).
 Notice that the low efficiency of the current-induced depinning (high $j_{p}/H_{p}$ ratio) 
 is not a disadvantage in itself. While the depinning current cannot be largely reduced
 with the magnetite nanostripe compared to that of the Py nanostripe, the depinning field can be
 and it is found smaller than the Walker-breakdown value in the magnetite nanostripes.
 This enables the field-only driven viscous motion of the $\pi$/2 DW in the presence
 of the notches. In the case of the $\pi$ DWs, such a motion was impossible
 in the Py nanostripes nor with the PMA nanostripes. 

We note that, for a head-to-head DW magnetized upward the nanostripe,
 the ratio of the depinning current and the depinning field for the nanostripe with
 the notch at upper edge (figure 4) is similar to that for the lower-edge patterned 
 nanostripe of figure 3. However, in the former case, the depinning current
 is only $j_{p}=0.44\cdot10^{12}{\rm Am}^{-2}$. 
 The strength of the mutual interactions of the head-to-head and tail-to-tail $\pi$/2 DWs
 requires a complex study (which is beyond the scope of the paper)
 in order to assess the minimal pinning field necessary. In particular, 
 the mediating role of the structure of the head-to-tail DWs (whose period is scalable
 with stripe thickness) is to be determined.

In order to determine the pace of the temperature increase due to the Joule heating, we apply
 the formula ${\rm d}T/{\rm d}t=j^{2}/\sigma C\rho$, where, $\sigma$ denotes the electrical
 conductivity, $C$ denotes the specific heat capacity, while $\rho$ denotes the mass density. 
 For Fe$_{3}$O$_{4}$ nanostripe, in the presence of the current of the density $j=10^{12}$A/m$^{2}$, 
 ($\sigma=1.0\cdot10^{4}$1/$\Omega$m, $C=0.67$J/gK, $\rho=5.0\cdot10^{3}$kg/m$^{3}$)
 \cite{zie07,naf13}, one estimates ${\rm d}T/{\rm d}t=3.0\cdot10^{4}$K/ns.
 The last value is significantly higher than for the Py nanostripe with the current 
 of the same density \cite{fan11,cou06}; ${\rm d}T/{\rm d}t=91$K/ns, which is due to small 
 electrical conductivity of the magnetite compared to Py. Notice that the Curie temperatures 
 of Py and Fe$_{3}$O$_{4}$ are similar, $T_{c}\approx850$K. Any hope for slowing down 
 the Joule heating of the magnetite nanostripe via material modifications (e.g. doping) in the 
 direction of reduction of the electrical resistivity is weak since Fe$_{3}$O$_{4}$ is the material
 of the lowest resistivity among all spinels. Hence, the limitation on the time length 
 of the current pulses usable to drive the DW propagation in the magnetite
 nanostripes ($\sim10$ps) is much stronger than for the Py stripes ($\sim1$ns).
 However, the spatial distances between the neighboring
 head-to-head and tail-to-tail $\pi/2$ DWs can be substantially reduced compared
 to the minimum spacing of the transverse/vortex DWs in the soft-magnetic nanostripes
 while the velocities of the DWs of the former type are an order of magnitude higher
 than those of the vortex DWs. Therefore, operating the Fe$_{3}$O$_{4}$ structures with ultra-short
 current pulses can be as efficient as operating the Py structures with the ns pulses.

\begin{figure}
\unitlength 1mm
\begin{center}
\begin{picture}(175,15)
\put(25,-5){\resizebox{87mm}{!}{\includegraphics{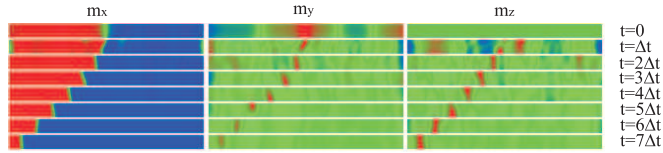}}}
\end{picture}
\end{center}
\caption{The stress-driven motion of the head-to-head DW in Fe$_{3}$O$_{4}$ nanostripe
of the $w\times\tau=100\times10$nm$^{2}$ cross-section. The stress components 
are $\sigma_{x}=0.6$GPa, $\sigma_{y}=2.4$GPa, $\sigma_{z}=0$.
The time step $\Delta t=0.5$ns.}
\end{figure} 

\begin{figure}
\unitlength 1mm
\begin{center}
\begin{picture}(175,18)
\put(25,-1){\resizebox{87mm}{!}{\includegraphics{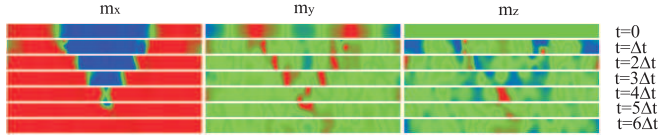}}}
\end{picture}
\end{center}
\caption{The mutual annihilation of two DWs (one head-to-head DW and one tail-to-tail DW)
via the stress-driven collision in Fe$_{3}$O$_{4}$ nanostripe of
the $w\times\tau=100\times10$nm$^{2}$ cross-section,
($\sigma_{x}=0.5$GPa, $\sigma_{y}=2.0$GPa, $\sigma_{z}=0$, $\Delta t=1$ns).
A transverse field of 30mT is applied in the direction (0,1,-1).}
\end{figure}  

\section{Stress-driven domain-wall motion}

We verify whether the stress application is an efficient alternative to the field
 application when driving of the DW propagation. With this purpose, we have performed
 the simulations of the stress-driven motion of the DW for the magnetite
 stripe of the width $w=100$nm and thickness $\tau=10nm$. The effects of the constant
 stress are easy to simulate with the MAGPAR package that we have utilized \cite{magpar}.
 The direction of the (in-the-plane) stress vector is taken with $\sigma_{x}/\sigma_{y}=1/4$,
 $\sigma_{z}=0$, where $x$ corresponds to the long axis of the stripe while $z$ axis
 coincides with the out-of-plane direction. The stripe length is 1.5$\mu$m. 
 
With the snapshots in figure 5,
 it is seen that the dynamics of the DW consists of two stages. Upon switching
 the stress on, the ferromagnetic domains undergo an almost instant reorientation because 
 the tetragonal magnetostriction creates novel easy directions. It forces
 the transition of the head-to-head (tail-to-tail) $\pi$/2 DW into the $\pi$-like DW 
 within 1ns approximately. In fact, the stress-induced anisotropy due to the tetragonal
 magnetostriction is two-axial while both easy (in-the-plane) directions are slightly deviated
 from the long axis of the stripe. At the second stage of the simulations, the internal field
 due to the rhombohedral magnetostriction drives the resulting $\pi$-like DW to propagate
 along the wire. The propagation takes place under sufficiently high total
 stress $\sigma\equiv|{\bf\sigma}|\ge0.6\sqrt{17}$GPa. 
 However, this lower bound on the motion-driving stress is not any fundamental value while
 the minimum stress changes with the simulation conditions (with the stripe length for instance).
 Its presence can be explained by changing 
 the preference of the system in terms of the stabilization of the $\pi$ DW or another 
 head-to-head DW (a $\pi-\epsilon\pi$ DW, where $\epsilon\ll1$) under the stress. The later
 structure propagates while the former is static as mentioned in Sec. II.  
 Turning on the stress in presence of a weak perpendicular magnetic field ensures
 the desired noncollinearity of the magnetization to be present.
 In particular, in the presence of a transverse in-the-plane field of 20mT, the simulated DW 
 propagates under the (subcritical) stress of $\sigma=0.5\sqrt{17}$GPa.

Notice that the stabilizing DWs are not in-the-plane magnetized while they are intermediate
 between the Bloch-like and Neel-like DWs. The stress-induced motion is almost viscous,
 however, minor oscillations of the DW structure are seen. 
   
The estimations based on the formulae of Appendix B, for the stress components
 $\sigma_{x}=0.6$GPa, $\sigma_{y}=2.4$GPa, $\sigma_{z}=0$ applied to the magnetite
 nanostripe 
%of $w\times t=100{\rm nm}\times10{\rm nm}$ 
(figure 5), lead to the DW width 
 $\Delta\approx26$nm for ${\rm cos}(\varphi)=1/10$ (an almost Bloch DW). 
 The velocity of the viscous propagation with dependence on the stress is given
 by formula (\ref{stress_mobility}). It leads,
 via $\mu_{\sigma}=270\cdot10^{3}$m/sGPa, 
 for the above values of the stress components, to the DW velocity 1600m/s that
 significantly overestimates the simulation result: 210m/s (in the relevant figure 5,
 the stripe length is 1.5$\mu$m). The discrepancy is not a surprise, however, since
 the DW motion is not fully turbulent (slow) nor purely viscous (fast), as seen from figure 5.
 The intermediate (oscillatory) character of the motion is expected to be accompanied by
 a decrease of the average velocity compared to the estimated value.

In figure 6, we demonstrate (by means of the micromagnetic simulations) the stress-induced
 collision of the head-to-head and tail-to-tail $\pi$/2 DWs and the possibility of annihilating
 the DW pairs. Because of an instability of the direction of the DW magnetization, a weak
 transverse (to the long axis) field of 30mT is applied and it is deviated from the stripe plane 
 by $\pi$/4 angle. Upon that stabilization, projections of the magnetizations of the colliding DWs
 are directed similar while the DW motion is viscous. Therefore, the expected result
 of the collision is mutual annihilation of the DWs following a general rule; the parallel
 (antiparallel) magnetized ferromagnetic DWs attract (repulse) \cite{jan13a,jan13b}.
 Our simulations confirm this expectation (figure 6).

Addressing the way of application of the stress to the nanostripe, we mention the strain-mediated   
 magnetoelectric effect in the multiferroic laminates. For instance, the magnetoelectricity
 of Fe$_{3}$O$_{4}$/PZN-PT and Fe$_{3}$O$_{4}$/PMN-PT is one of the strongest reported
 to date \cite{vaz12}. A rough estimation based on the data of reference \cite{liu10} shows sub-GPa
 in-the-plane stress to be induced in a quite thick (sub-$\mu$m) structural magnetite layer upon
 the application of the electric field of 6kV/cm in the PZN-PT substrate. Moreover, increasing 
 the stress in the magnetic layer via engineering the interface of the magnetoelectric composite
 is the subject of successive studies (e.g. \cite{gue14}). Therefore, we believe the desired values 
 of the stress to be achievable with a simple voltage application.

\section{Conclusions}

We have analyzed the field, current, and stress driven propagation of the $\pi$/2 DWs
 in the nanowires (nanostripes) created of cubic-anisotropy ferromagnetic materials. 
 with relevance to the potential application of those textures
 to the DW racetrack memory. The $\pi$/2 DWs are much narrower than the transverse DWs
 in the soft-magnetic
 nanowires of similar cross-sections which enables the recording with a bit density comparable 
 to that of the PMA nanowires. In spite of the small width of the $\pi$/2 DWs, the predicted
 velocities of the field-driven viscous motion of the $\pi$/2 DWs in the magnetite nanostripes 
 are higher than achievable for $\pi$ DWs in the Py nanostripes. It is because of large values
 of the critical field of the Walker breakdown in the magnetite nanostructures.
 Let us recall that such a viscous field-driven motion is not achievable with the single-layer
 PMA nanowires. 

The current-driven propagation
 of the $\pi$/2 DWs can be even faster than the fastest reported propagation of the chiral
 (Bloch) DWs in multi-layered nanowires with the PMA \cite{yan15}. 
 However, unlike for $\pi$ DW in the PMA nanowires, designing the racetrack for
 the $\pi$/2 DWs, there is no need to involve other mechanisms of the propagation driving
 with current than the direct STT. The conservation of the global chirality
 of the system of many $\pi$/2 DWs does not play any role in its positioning.

The possibility of the stress-driven propagation of the DWs arises due to the anisotropy 
 of the magnetostriction. We have demonstrated by means of the micromagnetic simulation
 the stress-induced collision of the DWs that results, under some stabilizing transverse field,
 in the annihilation of the DW pair. Hence, the stress application is found to be an alternative
 to using the longitudinal magnetic field when driving the domain magnetization reversal.

The promising cubic-anisotropy material considered is the magnetite.
 First, because of a relatively low saturation magnetization (a weak shape anisotropy) 
 and a strong cubic anisotropy, which enables the creation of the $\pi$/2 DWs.
 Second, because of large ratio of the rhombohedral magnetostriction constant
 to the tetragonal magnetostriction constant, which enables high-speed propagation of the DWs
 driven by the stress. Unfortunately, because of large electrical resistivity of the magnetite
 (the resistivity of the ferrites is two or three orders of magnitude larger than the resistivity
 of the Permalloy), the Joule heating remains a crucial problem in the racetrack design.
 However, small lengths of the domains and large velocities of the DWs enable
 as efficient current-driven positioning of the many-domain record as achievable with
 the Py-based racetracks, albeit using much shorter current pulses.

Finally, we notice that the material parameters of the Terfenol-D seem to make it
 the most relevant material to manufacture the racetrack of the $\pi$/2 DWs.
 It is of a very strong cubic anisotropy (of a negative value of the $K_{1}$ constant as desired) 
 and of an extremely strong rhombohedral magnetostriction.
 A relatively small saturation magnetization (a weak shape anisotropy
 compared to the crystalline anisotropy) leads to the formation of very narrow $\pi$/2 DWs
 in a wide range of thicknesses of the nanostripes (they are stable for up to 40nm-thick stripes).
 The electrical resistivity of the Terfenol-D is only $6.0\cdot10^{-5}\Omega$cm
 while the specific heat capacity and the mass density are comparable to those 
 of the magnetite and Py, thus, the Joule heating is slow \cite{coo00}.
 A quite big (estimated) value of the Gilbert damping constant of the Terfenol-D
 is a potential disadvantage that limits the DW velocity however. On the other hand,
 the brittleness of the Terfenol-D makes it difficult to process to the nanoscale, (while
 the nanotechnology of the magnetite is advanced \cite{elt08}).
     
\begin{appendix}		
\section{Field (current) driven dynamics of $\pi$/2 DW in 1D}

We treat a model of the $\pi$/2 DW motion analytically solving the LLG equation 
\begin{eqnarray}
\fl
-\frac{\partial\bi{m}}{\partial t}=\gamma\bi{m}\times\left(\bi{B}_{eff}
-\frac{\alpha}{M}\frac{\partial\bi{m}}{\partial t}
+\frac{\eta j}{\gamma M^{2}}\bi{m}\times\frac{\partial\bi{m}}{\partial x}
+\frac{\beta\eta j}{\gamma M}\frac{\partial\bi{m}}{\partial x}
\right)
\label{LLG}
\end{eqnarray}
in 1D. Here, $B_{eff}^{(i)}\equiv-\delta{\cal H}/\delta m_{i}$ while
\begin{eqnarray}
\fl
{\cal H}\equiv A\left(\frac{\partial\bi{m}}{\partial x}\right)^{2}
-\frac{K_{\|}}{M^{2}}(\bi{m}\cdot\hat{i})^{2}+\frac{K_{\bot}}{M^{2}}(\bi{m}\cdot\hat{k})^{2}
-\bi{m}\cdot\bi{B}+{\cal H}_{ca}+{\cal H}_{me}
\end{eqnarray}
Here, $\hat{i}\equiv(1,0,0)$, $\hat{k}\equiv(0,0,1)$, (the stripe is directed along the $x$ axis
 while $z$ axis is normal to the stripe plane), $\gamma$ denotes the gyromagnetic
 factor, $A$ denotes the exchange stiffness. The longitudinal external field $H$ is 
 defined via $\bi{B}=(\mu_{0}H,0,0)$. The effective easy-axis ($K_{\|}$)
 and easy-plane ($K_{\bot}$) anisotropies are of the shape (magnetostatic) origin. The 
 (cubic) crystalline anisotropy and the magnetoelastic coupling are included with 
 the ${\cal H}_{ca}$, ${\cal H}_{me}$ terms of the Hamiltonian, respectively. The constants 
 of the normal and anomalous STT are denoted with $\eta$ and $\eta\cdot\beta$,
 correspondingly, while $j$ denotes the density of the electric current. 
 
Parametrizing the magnetization with
 angles; $\bi{m}=M[\cos(\theta),\sin(\theta)\cos(\varphi),\sin(\theta)\sin(\varphi)]$, one describes
 a $\pi/2$ DW texture with 
 $\theta=\pi/4+{\rm arctan}(b/a)$, $b=1$, $a={\rm exp}\{[x-q(t)]/\Delta(t)\}$.
% In fact, the domain magnetization in the cubic-anisotropy nanostripes is a bit deviated 
% from the $(\theta=\pm\pi/4,\varphi=0)$ directions onto the long axis of the stripe
% because of the shape anisotropy, which is included via the tilting parameter $0<\epsilon\\1$.
 That ansatz for solving the equation (\ref{LLG}) is a modified (via including the dependence
 of $q$ and $\Delta$ on the time) static solution to the LLG equation \cite{jan14}.
 Inserting the ansatz into (\ref{LLG}), we notice that
 $a/\sqrt{(a^{2}+b^{2})}=\cos(\theta-\pi/4)$, $b/\sqrt{(a^{2}+b^{2})}=\sin(\theta-\pi/4)$, thus,
 $\sin(\theta)=(a+b)/\sqrt{2(a^{2}+b^{2})}$,
 $\cos(\theta)=(a-b)/\sqrt{2(a^{2}+b^{2})}$.
 For $\varphi=0$ (the in-the-plane ordering), and $\dot{\Delta}=0$ (a sufficiently-long
 time upon switching the field or current on \cite{zha04,tse08}), one arrives at the system
 of the equations
\begin{eqnarray}
\fl
\frac{1}{\Delta}\left(\dot{q}+\eta j\right)\frac{ab}{\sqrt{a^{2}+b^{2}}}
+\alpha\dot{\varphi}\frac{(a+b)}{\sqrt{2}}=0,\nonumber\\
\fl
\left(\frac{2\gamma A}{\Delta^{2}M}+\frac{2\gamma K_{1}}{M}\right)
\frac{ab(b^{2}-a^{2})}{(a^{2}+b^{2})^{3/2}}
-\frac{\gamma K_{\|}}{M}\frac{b^{2}-a^{2}}{\sqrt{a^{2}+b^{2}}}\label{rough}\\
-\left(\gamma\mu_{0}H+\dot{\varphi}\right)\frac{(a+b)}{\sqrt{2}}
+\left(\frac{\alpha}{\Delta}\dot{q}+\frac{\beta\eta j}{\Delta}\right)
\frac{ab}{\sqrt{a^{2}+b^{2}}}&=&0.\nonumber
\end{eqnarray}
In the center of the DW [for $x=q(t)$], thus, for $a=b=1$, these equations simplify to
\begin{eqnarray} 
%\Delta=\sqrt{\frac{A}{K_{\|}-K_{1}}},\nonumber\\
\frac{\dot{q}+\eta j}{2\Delta}+\alpha\dot{\varphi}=0,\nonumber\\
\gamma\mu_{0}H+\dot{\varphi}-\frac{\alpha\dot{q}+\beta\eta j}{2\Delta}=0.
\label{reduced_eq}
\end{eqnarray}
A rough estimate of the width of the DW using (\ref{rough}) leads
 to $\Delta=\sqrt{A/(K_{\|}-K_{1})}$, where $|K_{1}|\gg 1/10\mu_{0}M^2>K_{\|}$
 for the aspect ratio $w/t\approx10$, (estimation details follow reference \cite{bea08}).

The first of equations (\ref{rough}) and the first of equations (\ref{reduced_eq}) become irrelevant
 to the description of the viscous DW motion since they originate from the dynamical equation 
 of the out-of-plane magnetization component $m_{z}$. One removes them \cite{thi05}, thus,
 finding $\dot{q}=(-\beta\eta j+\gamma\Delta H)/\alpha$.

\section{Stress-driven dynamics of DW in 1D}

We consider the head-to-head or tail-to-tail DW under an in-the-plane directed 
 and sufficiently-strong external stress. According to the simulation of section IV,
 the domains which are magnetized in the $(1,\pm1,0)$ directions in the absence
 of the stress rapidly remagnetize onto the long axis of the stripe upon switching
 the stress on. Thus, the head-to-tail DWs disappear while the head-to-head
 (tail-to-tail) $\pi$/2 DWs transform into $\pi$ DWs. The domain remagnetization
 is not complete however, while the magnetization direction
 is weakly deviated from the longitudinal one. That enables the creation of the 
 longitudinal field inside the domains with the stress, thus, driving the DW motion can
 be explained with minimizing the energy via shrinking (developing) the domains magnetized
 almost antiparallel (parallel) to the field (similar to minimizing the Zeeman energy via the DW
 propagation). Additionally, an in-the-plane anisotropy is created with the stress, which
 influences the DW width. 

In order to model the stress-induced dynamics of the $\pi-\epsilon$ DW in the limit
 $\epsilon\to0^{+}$, we apply the standard ansatz 
 $\theta=2{\rm arctan}(b/a)$, $b=1$, $a={\rm exp}\{[x-q(t)]/\Delta(t)\}$
 that is used to describe the motion of the $\pi$ DW in the wires. 
 Notice that $\sin(\theta)=2ab/(a^{2}+b^{2})$, $\cos(\theta)=(a^{2}-b^{2})/(a^{2}+b^{2})$.
 We consider the long-term dynamics, $\dot{\Delta}=0$,
 in the absence of the external magnetic field and electric current.
 The substitution of the ansatz into the LLG system, for $|\varphi|\ll\pi/2$, leads to
\begin{eqnarray}
\fl
\frac{\dot{q}}{\Delta}+\alpha\dot{\varphi}=0,\nonumber\\
\fl
-\dot{\varphi}+\frac{\alpha}{\Delta}\dot{q}
+\left[-\frac{2\gamma[K_{\|}+K_{\bot}{\rm sin}^{2}(\varphi)]}{M}
-\frac{2\gamma K_{1}}{M}\frac{(a^{2}-b^{2})^{2}-4a^{2}b^{2}}{(a^{2}+b^{2})^{2}}
+\frac{2\gamma A}{\Delta^{2}M}
\right.
\label{stress_equations}\\\fl\left.
+\frac{3\gamma\lambda_{100}\sigma{\rm cos}(\varphi)}{M}(\gamma_{2}^{2}-\gamma_{1}^{2})\right]
\frac{b^{2}-a^{2}}{a^{2}+b^{2}}
-\frac{3\gamma\lambda_{111}\sigma\gamma_{1}\gamma_{2}}{M}
\frac{(a^{2}-b^{2})^{2}-4a^{2}b^{2}{\rm cos}^{2}(\varphi)}{2ab(a^{2}+b^{2})}=0,
\nonumber
\end{eqnarray}
where $\gamma_{1}\equiv\sigma_{x}/\sigma$, $\gamma_{2}\equiv\sigma_{y}/\sigma$, and
 $\sigma\equiv|{\bf\sigma}|$. In the center of the DW, $a=b=1$.
Assuming the range of stresses 
$K_{\|}+K_{\bot}{\rm sin}^{2}(\varphi)\ll-K_{1}-3\lambda_{100}\sigma(\gamma_{2}^{2}-\gamma_{1}^{2}){\rm cos}(\varphi)$,
the second equation of (\ref{stress_equations}) leads to
\begin{eqnarray}
\Delta=\sqrt{\frac{2A}{-2K_{1}-3\lambda_{100}\sigma(\gamma_{2}^{2}-\gamma_{1}^{2}){\rm cos}(\varphi)}}.
\end{eqnarray}
The last relation limits the range of $\varphi$ angle to ${\rm cos}(\varphi)>0$, thus,
 the rotation of the DW magnetization, if takes place, is not stationary while oscillatory.  
 We evaluate the velocity of the stress-driven (viscous) motion of the DW  
\begin{eqnarray}
\dot{q}=\mu_{\sigma}{\rm cos}^{2}(\varphi)\gamma_{1}\gamma_{2}\sigma,
\label{stress_mobility}\\
\mu_{\sigma}\equiv\frac{3\gamma\lambda_{111}\Delta}{M\alpha}.
\nonumber
\end{eqnarray}
Following (\ref{stress_mobility}),
 since $\pi$ DWs are completely out-of-the-plane magnetized, ($\varphi=\pm\pi/2$ in the 
 center of the $\pi$ DW), they cannot be driven by the stress, whereas, $\pi-\epsilon$ DWs
 which separate slightly noncollinear domains can, ($\varphi\neq\pm\pi/2$ in the center of
 such a DW).
\end{appendix}

\end{document}